\documentclass[a4paper,USenglish,cleveref]{lipics-v2019}
\title{Linear Time Subgraph Counting, Graph Degeneracy, and the Chasm at Size Six}

\author{Suman K. Bera}{University of California, Santa Cruz, CA 95064, USA}{sbera@ucsc.edu}{}{}

\author{Noujan Pashanasangi}{University of California, Santa Cruz, CA 95064, USA}{npashana@ucsc.edu}{}{}

\author{C. Seshadhri}{University of California, Santa Cruz, CA 95064, USA}{sesh@ucsc.edu}{}{}

\authorrunning{S. K. Bera, N. Pashanasangi, C. Seshadhri}%TODO mandatory. First: Use abbreviated first/middle names. Second (only in severe cases): Use first author plus 'et al.'

\Copyright{Suman K. Bera, Noujan Pashanasangi, and C. Seshadhri}



\funding{All authors are supported by NSF TRIPODS grant CCF-1740850 NSF CCF-1813165, and ARO Award W911NF1910294}

\EventEditors{Thomas Vidick}
\EventNoEds{1}
\EventLongTitle{11th Innovations in Theoretical Computer Science Conference (ITCS 2020)}
\EventShortTitle{ITCS 2020}
\EventAcronym{ITCS}
\EventYear{2020}
\EventDate{January 12--14, 2020}
\EventLocation{Seattle, Washington, USA}
\EventLogo{}
\SeriesVolume{151}
\ArticleNo{38}

\usepackage{xspace,verbatim,bm,bbm}
\usepackage{multicol,booktabs}
\usepackage{algorithm,algorithmicx}
\usepackage[noend]{algpseudocode}

\usepackage{url}
\usepackage{esvect}

\usepackage[usenames, dvipsnames]{xcolor}

\usepackage{mathdots,mathtools,mathrsfs}
\usepackage{tikz}

\nolinenumbers

\definecolor{airforceblue}{rgb}{0.36, 0.54, 0.66}
\definecolor{bluegray}{rgb}{0.4, 0.6, 0.8}
\definecolor{ceil}{rgb}{0.57, 0.63, 0.81}
\definecolor{celestialblue}{rgb}{0.29, 0.59, 0.82}
\definecolor{cerulean}{rgb}{0.0, 0.48, 0.65}
\definecolor{celadon}{rgb}{0.67, 0.88, 0.69}
\definecolor{yellow-green}{rgb}{0.6, 0.8, 0.2}

\colorlet{graph-vertex-blue}{blue}
\colorlet{graph-vertex-green}{Green}
\colorlet{graph-vertex-red}{red}
\colorlet{graph-vertex-yellow}{yellow}

\colorlet{graph-edge-blue}{blue}
\colorlet{graph-edge-green}{Green}
\colorlet{graph-edge-red}{red}
\colorlet{graph-edge-yellow}{yellow}

\usepackage{thmtools}
\usepackage{thm-restate}

\newtheorem{observation}[theorem]{Observation}
\theoremstyle{definition}
\newtheorem{conjecture}[theorem]{Conjecture}

\usepackage[normalem]{ulem}

\newcommand{\ignore}[1]{}

\newcommand{\etal}{{et al.}\xspace}

\newcommand{\mcA}{\mathcal A}

\newcommand{\mcC}{\mathcal C}

\newcommand{\mcH}{\mathcal H}

\newcommand{\mcK}{\mathcal K}

\newcommand{\mcM}{\mathcal M}

\newcommand{\mcT}{\mathcal T}

\usepackage{hyperref}
\newcommand{\Sec}[1]{\hyperref[sec:#1]{\S\ref*{sec:#1}}} %section
\newcommand{\Eqn}[1]{\hyperref[eq:#1]{(\ref*{eq:#1})}} %equation
\newcommand{\Fig}[1]{\hyperref[fig:#1]{Fig.\,\ref*{fig:#1}}} %figure
\newcommand{\Tab}[1]{\hyperref[tab:#1]{Tab.\,\ref*{tab:#1}}} %table
\newcommand{\Thm}[1]{\hyperref[thm:#1]{Theorem\,\ref*{thm:#1}}} %theorem
\newcommand{\Fact}[1]{\hyperref[fact:#1]{Fact\,\ref*{fact:#1}}} %fact
\newcommand{\Lem}[1]{\hyperref[lem:#1]{Lemma\,\ref*{lem:#1}}} %lemma
\newcommand{\Prop}[1]{\hyperref[prop:#1]{Prop.~\ref*{prop:#1}}} %property
\newcommand{\Cor}[1]{\hyperref[cor:#1]{Corollary~\ref*{cor:#1}}} %corollary
\newcommand{\Conj}[1]{\hyperref[conj:#1]{Conjecture~\ref*{conj:#1}}} %conjecture
\newcommand{\Def}[1]{\hyperref[def:#1]{Definition~\ref*{def:#1}}} %definition
\newcommand{\Alg}[1]{\hyperref[alg:#1]{Alg.~\ref*{alg:#1}}} %algorithm
\newcommand{\Ex}[1]{\hyperref[ex:#1]{Ex.~\ref*{ex:#1}}} %example
\newcommand{\Clm}[1]{\hyperref[clm:#1]{Claim~\ref*{clm:#1}}} %example

\renewcommand{\ge}{\geqslant}
\renewcommand{\le}{\leqslant}
\renewcommand{\geq}{\geqslant}
\renewcommand{\leq}{\leqslant}

\newcommand{\before}{\mathrel{\lhd}}

\DeclareMathOperator{\mm}{max}

\newcommand{\degen}{\kappa}
\newcommand{\arbo}{\alpha}

\newcommand{\SUBC}{\textsc{sub-cnt}\xspace}
\newcommand{\TRIC}{\textsc{tri-cnt}\xspace}
\newcommand{\TRICONJ}{\textsc{Triangle Detection Conjecture}\xspace}
\newcommand{\threesum}{\textsc{3SUM Conjecture}\xspace}

\newcommand{\matchCount}{\hbox{M}}

\DeclareMathOperator{\matchCountDist}{DM}
\newcommand{\hashmap}{\mcH\mcM}

\newcommand{\dir}{^\rightarrow}

\newcommand{\dgnord}{^{\rightarrow}_{\mathrel{\lhd}}}
\newcommand{\drtss}{\hbox{DRTS}}

\usepackage{tikz}
\usetikzlibrary{decorations.markings,fit,arrows,positioning,backgrounds}
\tikzset{%
  gnode/.style={shape=circle,minimum size=3mm,fill,draw=black}
}
\tikzset{myptr/.style={decoration={markings,mark=at position 1 with {\arrow[scale=1.5,>=stealth]{>}}},postaction={decorate}}}

\tikzset{myptr2/.style={decoration={markings,mark=at position 0.55 with {\arrow[scale=1.5,>=stealth]{>}}},postaction={decorate}}}

\def\CountableSixVertexExample{
   \node (1) at (0.9,0) [nd] {};
   \node (2) at (0,0.75) [nd] {};
   \node (3) at (0,1.5) [nd] {};
   \node (4) at (0.9,2.25) [nd] {};
   \node (5) at (1.8,1.5) [nd] {};
   \node (6) at (1.8,0.75) [nd] {};
   \draw (1) to (2);
   \draw (1) to (6);
   \draw (2) to (3);
   \draw (2) to (5);
   \draw (3) to (6);
   \draw (5) to (6);
   \draw (4) to (3);
   \draw (4) to (5);
   \draw (4) to (1);
   
   \node (7) [inner sep = 0] at (2.5,1.25) {};
   \node (8) [inner sep = 0] at (3.3,1.25) {};
   \draw[myptr, color=black!30!green] (7)  to (8);
   
   \node (9) at (4.9,0) [nd, fill = graph-vertex-green] {};
   \node (10) at (4,0.75) [nd] {};
   \node (11) at (4,1.5) [nd, fill = graph-vertex-green] {};
   \node (12) [label={above:\small $s$}] at (4.9,2.25) [nd, fill = graph-vertex-green] {};
   \node (13) at (5.8,1.5) [nd, fill = graph-vertex-green] {};
   \node (14) at (5.8,0.75) [nd] {};
   \draw (9) to (10);
   \draw (9) to (14);
   \draw (10) to (13);
   \draw (10) to (11);
   \draw (11) to (14);
   \draw (13) to (14);
   \draw[myptr, color=graph-edge-green] (12) to (11);
   \draw[myptr, color=graph-edge-green] (12) to (13);
   \draw[myptr, color=graph-edge-green] (12) to (9);
   
  \node [inner sep = 0,below] at +(1,-0.6) {{\Large $H$}};
  \node [inner sep = 0,below] at +(5,-0.6) {{\Large $H\dir$}};
}

\def\UncountableSixVertexExample{
   \node (1) at (0.9,0) [nd] {};
   \node (2) at (0,0.75) [nd] {};
   \node (3) at (0,1.5) [nd] {};
   \node (4) at (0.9,2.25) [nd] {};
   \node (5) at (1.8,1.5) [nd] {};
   \node (6) at (1.8,0.75) [nd] {};
   \draw (1) to (2);
   \draw (1) to (6);
   \draw (3) to (2);
   \draw (3) to (4);
   \draw (5) to (4);
   \draw (5) to (6);
   
   \node (7) [inner sep = 0] at (2.5,1.25) {};
   \node (8) [inner sep = 0] at (3.3,1.25) {};
   \draw[myptr, color=black!30!red] (7)  to (8);

   \node (9)  [label={below:\small $s_1$}] at (4.9,0) [nd, fill = graph-vertex-red] {};
   \node (10) [label={left: \small $t_1$}] at (4,0.75) [nd, fill = graph-vertex-red] {};
   \node (11) [label={left: \small $s_2$}]  at (4,1.5) [nd] {};
   \node (12) [label={above:\small $t_2$}] at (4.9,2.25) [nd] {};
   \node (13) [label={right:\small $s_3$}] at (5.8,1.5) [nd] {};
   \node (14) [label={right:\small $t_3$}] at (5.8,0.75) [nd, fill = graph-vertex-red] {};
   \draw[myptr, color=graph-edge-red] (9) to (10);
   \draw[myptr, color=graph-edge-red] (9) to (14);
   \draw[myptr] (11) to (10);
   \draw[myptr] (11) to (12);
   \draw[myptr] (13) to (12);
   \draw[myptr] (13) to (14);
   
   \node [inner sep = 0,below] at +(1,-0.6) {{\Large $H$}};
   \node [inner sep = 0,below] at +(5,-0.6) {{\Large $H\dir$}};
}

\def\FiveVertexExample{
   \node (1) [label = {below:\small $e$}] at (2,0) [nd] {};
   \node (2) [label = {left:\small $a$}] at (0.3,1) [nd] {};
   \node (3) [label = {right:\small $b$}] at (0.9,1) [nd] {};
   \node (4) [label = {right:\small $c$}] at (2.8,1) [nd] {};
   \node (5) [label = {above:\small $d$}] at (2,2) [nd] {};
   
   \draw (1) to (2);
   \draw (1) to (3);
   \draw (1) to (4);
   \draw (1) to (5);
   \draw (2) to (5);
   \draw (3) to (5);
   \draw (4) to (5);
   
   \node [inner sep = 0,below] at +(2,-0.6) {{\Large $H$}};
}

\def\DirFiveVertexExample{
   \node (1) [label = {below:\small $e$}] at (2,0) [nd, fill=graph-vertex-green] {};
   \node (2) [label = {left:\small $a$}] at (0.3,1) [nd] {};
   \node (3) [label = {right:\small $b$}] at (0.9,1) [nd] {};
   \node (4) [label = {right:\small $c$}] at (2.8,1) [nd, fill=graph-vertex-green] {};
   \node (5) [label = {above:\small $d$}] at (2,2) [nd, fill=graph-vertex-green] {};
   
   \draw[myptr2] (2) to (1);
   \draw[myptr2] (3) to (1);
   \draw[myptr2, color=graph-edge-green] (4) to (1);
   \draw[myptr2] (5) to (1);
   \draw[myptr2] (2) to (5);
   \draw[myptr2] (3) to (5);
   \draw[myptr2, color=graph-edge-green] (4) to (5);
   
   \node [inner sep = 0,below] at +(2,-0.6) {{\Large $H\dir$}};
}

\def\DirFiveVertexExampleOneMatch{
   \node (1) [label = {below:\small $y:e$}] at (2,0) [nd, fill=graph-vertex-green] {};
   \node (2) [label = {left:\small $u:a$}] at (0.3,1) [nd] {};
   \node (3) [label = {right:\small $v:b$}] at (0.9,1) [nd] {};
   \node (4) [label = {right:\small $w:c$}] at (2.8,1) [nd, fill=graph-vertex-green] {};
   \node (5) [label = {above:\small $x:d$}] at (2,2) [nd, fill=graph-vertex-green] {};
   
   \draw[myptr2] (2) to (1);
   \draw[myptr2] (3) to (1);
   \draw[myptr2, color=graph-edge-green] (4) to (1);
   \draw[myptr2] (5) to (1);
   \draw[myptr2] (2) to (5);
   \draw[myptr2] (3) to (5);
   \draw[myptr2, color=graph-edge-green] (4) to (5);
   
   \node [inner sep = 0,below] at +(2,-0.8) {{\Large $\pi_1$}};
}

\def\DirFiveVertexExampleTwoMatch{
   \node (1) [label = {below:\small $y:e$}] at (2,0) [nd, fill=graph-vertex-green] {};
   \node (2) [label = {left:\small $u:b$}] at (0.3,1) [nd] {};
   \node (3) [label = {right:\small $v:a$}] at (0.9,1) [nd] {};
   \node (4) [label = {right:\small $w:c$}] at (2.8,1) [nd, fill=graph-vertex-green] {};
   \node (5) [label = {above:\small $x:d$}] at (2,2) [nd, fill=graph-vertex-green] {};

   \draw[myptr2] (2) to (1);
   \draw[myptr2] (3) to (1);
   \draw[myptr2, color=graph-edge-green] (4) to (1);
   \draw[myptr2] (5) to (1);
   \draw[myptr2] (2) to (5);
   \draw[myptr2] (3) to (5);
   \draw[myptr2, color=graph-edge-green] (4) to (5);
   
   \node [inner sep = 0,below] at +(2,-0.8) {{\Large $\pi_2$}};
}

\def\DirFiveVertexExampleThreeMatch{
   \node (1) [label = {below:\small $y:e$}] at (2,0) [nd, fill=graph-vertex-green] {};
   \node (2) [label = {left:\small $u:a$}] at (0.3,1) [nd] {};
   \node (3) [label = {right:\small $v:c$}] at (0.9,1) [nd, fill=graph-vertex-green] {};
   \node (4) [label = {right:\small $w:b$}] at (2.8,1) [nd] {};
   \node (5) [label = {above:\small $x:d$}] at (2,2) [nd, fill=graph-vertex-green] {};
   
   \draw[myptr2] (2) to (1);
   \draw[myptr2, color=graph-edge-green] (3) to (1);
   \draw[myptr2] (4) to (1);
   \draw[myptr2] (5) to (1);
   \draw[myptr2] (2) to (5);
   \draw[myptr2, color=graph-edge-green] (3) to (5);
   \draw[myptr2] (4) to (5);
   
   \node [inner sep = 0,below] at +(2,-0.8) {{\Large $\pi_3$}};
}

\def\DirFiveVertexExampleFourMatch{
   \node (1) [label = {below:\small $y:e$}] at (2,0) [nd, fill=graph-vertex-green] {};
   \node (2) [label = {left:\small $u:b$}] at (0.3,1) [nd] {};
   \node (3) [label = {right:\small $v:c$}] at (0.9,1) [nd, fill=graph-vertex-green] {};
   \node (4) [label = {right:\small $w:a$}] at (2.8,1) [nd] {};
   \node (5) [label = {above:\small $x:d$}] at (2,2) [nd, fill=graph-vertex-green] {};
   
   \draw[myptr2] (2) to (1);
   \draw[myptr2, color=graph-edge-green] (3) to (1);
   \draw[myptr2] (4) to (1);
   \draw[myptr2] (5) to (1);
   \draw[myptr2] (2) to (5);
   \draw[myptr2, color=graph-edge-green] (3) to (5);
   \draw[myptr2] (4) to (5);
   
   \node [inner sep = 0,below] at +(2,-0.8) {{\Large $\pi_4$}};
}

\def\DirFiveVertexExampleFiveMatch{
   \node (1) [label = {below:\small $y:e$}] at (2,0) [nd, fill=graph-vertex-green] {};
   \node (2) [label = {left:\small $u:c$}] at (0.3,1) [nd, fill=graph-vertex-green] {};
   \node (3) [label = {right:\small $v:a$}] at (0.9,1) [nd] {};
   \node (4) [label = {right:\small $w:b$}] at (2.8,1) [nd] {};
   \node (5) [label = {above:\small $x:d$}] at (2,2) [nd, fill=graph-vertex-green] {};   
   
   \draw[myptr2, color=graph-edge-green] (2) to (1);
   \draw[myptr2] (3) to (1);
   \draw[myptr2] (4) to (1);
   \draw[myptr2] (5) to (1);
   \draw[myptr2, color=graph-edge-green] (2) to (5);
   \draw[myptr2] (3) to (5);
   \draw[myptr2] (4) to (5);
   
   \node [inner sep = 0,below] at +(2,-0.8) {{\Large $\pi_5$}};
}

\def\DirFiveVertexExampleSixMatch{
   \node (1) [label = {below:\small $y:e$}] at (2,0) [nd, fill=graph-vertex-green] {};
   \node (2) [label = {left:\small $u:c$}] at (0.3,1) [nd, fill=graph-vertex-green] {};
   \node (3) [label = {right:\small $v:b$}] at (0.9,1) [nd] {};
   \node (4) [label = {right:\small $w:a$}] at (2.8,1) [nd] {};
   \node (5) [label = {above:\small $x:d$}] at (2,2) [nd, fill=graph-vertex-green] {};   
   
   \draw[myptr2, color=graph-edge-green] (2) to (1);
   \draw[myptr2] (3) to (1);
   \draw[myptr2] (4) to (1);
   \draw[myptr2] (5) to (1);
   \draw[myptr2, color=graph-edge-green] (2) to (5);
   \draw[myptr2] (3) to (5);
   \draw[myptr2] (4) to (5);
   
   \node [inner sep = 0,below] at +(2,-0.8) {{\Large $\pi_6$}};
}

\def\EdgeSplitSixCycle{
   \node (1) [label = {below:\small $u$}] at (0.3,1) [nd] {};
   \node (2) [label = {below:\small $v$}] at (1.8,1) [nd] {};
   
   \draw (1) to  node[above, midway] {$e$} (2);
   
   \node (3) [inner sep = 0] at (2.4,1) {};
   \node (4) [inner sep = 0] at (3.4,1) {};
   \draw[myptr, color=black] (3)  to (4);
   
   \node (5) [label = {below:\small $u$}] at (4,1) [nd] {};
   \node (6) [label = {above:\small $v_e$}] at (5.2,1.5) [nd,fill=graph-vertex-red] {};
   \node (7) [label = {below:\small $v$}] at (6.4,1) [nd] {};

   \draw (5) to node[above, midway] {$e_1$} (6);
   \draw (6) to node[above, midway] {$e_2$} (7);

   \node [inner sep = 0,below] at +(1.2,-1) {{\Large $e\in E(G)$}};
   \node [inner sep = 0,below] at +(5.3,-1) {{\Large $e_1,e_2 \in E(G_6)$}};
}

\def\EdgeSplitSevenCycle{
   \node (1) [label = {below:\small $u$}] at (0.3,1) [nd] {};
   \node (2) [label = {below:\small $v$}] at (1.8,1) [nd] {};
   
   \draw (1) to  node[above, midway] {$e$} (2);
   
   \node (3) [inner sep = 0] at (2.4,1) {};
   \node (4) [inner sep = 0] at (3.4,1) {};
   \draw[myptr, color=black] (3)  to (4);
   
   \node (5) [label = {left:\small $u$}] at (4,1) [nd] {};
   \node (6) [label = {above:\small $v_{e}$}] at (5.5,2) [nd, fill=graph-vertex-red] {};
   \node (7) [label = {right:\small $v$}] at (7,1) [nd] {};
   \node (8) [label = {left:\small $u_{e_1}$}] at (4.75,0) [nd, fill=graph-vertex-red] {};
   \node (9) [label = {right:\small $u_{e_2}$}] at (6.25,0) [nd,fill=graph-vertex-red] {};

   \draw (5) to node[above, near start] {$e_{1,1}$} (6);
   \draw (6) to node[above, near end] {$e_{1,2}$} (7);
   \draw (5) to node[left, midway] {$e_{2,1}$} (8);
   \draw (8) to node[below, midway] {$e_{2,2}$} (9);
   \draw (9) to node[right, midway] {$e_{2,3}$} (7);   

   \node [inner sep = 0,below] at +(1.2,-1) {{\Large $e\in E(G)$}};
   \node [inner sep = 0,below] at +(5.3,-1) {{\Large $e_{i,j}  \in E(G_7)$}};
}

\hideLIPIcs

\begin{document}

\maketitle

\begin{abstract}
We consider the problem of counting all $k$-vertex
subgraphs in an input graph, for any constant $k$.
This problem (denoted $\SUBC_k$) has been studied extensively in
both theory and practice. In a classic result, Chiba and Nishizeki (SICOMP 85)
gave linear time algorithms for clique and 4-cycle counting
for \emph{bounded degeneracy graphs}. This is a rich class of sparse graphs
that contains, for example, all minor-free families and preferential attachment
graphs. The techniques from this result have
inspired a number of recent practical algorithms for $\SUBC_k$. 
Towards a better understanding of the limits of these techniques, we ask:
for what values of $k$ can $\SUBC_k$ be solved in linear time? 

We discover a chasm at $k=6$. Specifically, we prove that for $k < 6$, $\SUBC_k$
can be solved in linear time. Assuming a standard conjecture in fine-grained
complexity, we prove that for all $k \geq 6$, $\SUBC_k$ cannot be solved 
even in near-linear time.
\end{abstract}

\section{Introduction}

The subgraph counting problem asks for the number of occurrences of a (typically connected)
``pattern'' subgraph $H$ in a connected input graph $G$. It is a fundamental algorithmic problem
with a rich theory~\cite{itai1978finding,monien1985find,chiba1985arboricity,nevsetvril1985complexity,alon1997finding,eisenbrand2004complexity,vassilevska2009efficient,curticapean2017homomorphisms}, and widely used in practice~\cite{HoLe70,Co88,Po98,Burt04,PrzuljCJ04,HoBe+07, Pr07,khan2011neighborhood,UganderBK13,SaSe15,Ts15,BeGlLe16}. With the explosion of network science, subgraph counting is now a fundamental tool used for analyzing
real-world graphs. Thus, the search for fast algorithms for subgraph counting is not just a theoretical
problem, but one that has many applications in bioinformatics, social sciences,
and computer science. 

Especially for the many of the practical
applications, a common version of subgraph counting is to count the frequency
of \emph{all connected subgraphs} with $k$ vertices~\cite{PiSeVi17,ortmann2017efficient,AhNe+15,elenberg2015beyond,elenberg2016distributed,hovcevar2016computation,marcus2010efficient,RaBhHa14,jha2015path,wang2017moss,wang2017moss}.
We will denote this problem as $\SUBC_k$. Even in the theory literature,
it is common to parametrize running time by $n$ (vertices in $G$) and $k$,
so it is natural to study $\SUBC_k$.
There is a rich line of theoretical work on getting $n^{\mu k}$ time algorithms, for $\mu < 1$,
using matrix multiplication and tree decomposition methods~\cite{itai1978finding,nevsetvril1985complexity,alon1997finding,bjorklund2009counting,koutis2009limits,vassilevska2009finding,kowaluk2013counting,curticapean2014complexity,bjorklund2017counting,curticapean2017homomorphisms}. Unfortunately,
$\SUBC_k$ is (a generalization of) the canonical \#W[1]-hard problem, and it
is not believed that there exist $f(k)\cdot n^{o(k)}$ algorithms for $\SUBC_k$.
From an application standpoint, these algorithms are typically not practical, and do not provide algorithmic guidance. 
Real-world graphs are massive, and one typically desires linear-time algorithms.

An alternate perspective is to look for faster algorithms for restricted graph
classes, and hope that these classes correspond to real-world graphs.
A seminal result of Chiba-Nishizeki gave $O(m\degen^{k-2})$ algorithms for $k$-clique
counting and an $O(m\degen)$ algorithm for $4$-cycle counting, where $m$ is the number of edges in $G$ and $\degen$ is the \emph{graph
degeneracy}~\cite{chiba1985arboricity}. We leave the technical definitions for later; but $\degen$
can be thought of as the maximum average degree of any subgraph of $G$. 
Chiba-Nishizeki implicitly prove 
linear-time algorithms for $\SUBC_k$ for $k = 3, 4$ (explicitly shown in~\cite{PiSeVi17,ortmann2017efficient}
).
The class of bounded (constant) degeneracy graphs is immensely rich:
it contains all minor-closed families, preferential attachment graphs,
and bounded expansion graphs. 
The graph degeneracy appears heavily in network science, and real-world
graphs have typically low degeneracy (though maybe not constant).

But most importantly for subgraph counting, the techniques from Chiba-Nishizeki
have inspired a number of recent practical subgraph counting algorithms~\cite{PiSeVi17,ortmann2017efficient,jha2015path,jain2017fast}.

The problems of $\SUBC_k$ for $k \leq 5$ have been successfully tackled in practice using these approaches. These algorithms are often tailored for $k$ (using,
for example, specific tricks to count individual $4$-vertex subgraphs) and it
is not clear how far they will extend for larger $k$.

Towards a better theoretical understanding, we pose the following question.

\vspace{0.2cm}
\emph{For what values $k$, does
the $\SUBC_k$ problem admit a linear time algorithm in bounded degeneracy graphs?}

\subsection{Our Results}

The question above has a surprisingly clean resolution, assuming
conjectures from fine-grained complexity. For simplicity,
we assume that the input graph $G$ is connected. We assume Las Vegas randomized algorithms,
so we talk of expected running times.

Our main theorem asserts linear time algorithms for counting (up to) 5-vertex
subgraphs in bounded degeneracy graphs. For counting 6-vertex subgraphs and beyond, 
it is unlikely that even near-linear time algorithms exists.

\begin{theorem} [The chasm at size $6$] \label{thm:informal} For $k \leq 5$, there is an expected  $O(m\degen^{k-2})$ time  algorithm for $\SUBC_k$.

Assume the \TRICONJ (Conj. \ref{conj:triangle}). There exists an absolute constant $\gamma > 0$ such that the following
holds. For any $k \geq 6$ and any function $f:\mathbb{N} \to \mathbb{N}$, there
is no (expected) $o(m^{1+\gamma}f(\degen))$ algorithm for $\SUBC_k$.
\end{theorem}

The \TRICONJ was first  stated by Abboud and Williams~\cite{abboud2014popular}.
They proved many
lower bounds for the dynamic version of many well known graph problems such as 
bipartite perfect matching, single source reachability etc. It is actually believed
that the constant $\gamma$ could be as large as $1/3$.

\begin{conjecture}[\TRICONJ~\cite{abboud2014popular}] 
\label{conj:triangle}
There exists a constant $\gamma>0$ such that in the word RAM model
of $O(\log n)$ bits, any algorithm to detect whether an input graph on
$m$ edges has a triangle requires $\Omega(m^{1+\gamma})$ time
in expectation.
\end{conjecture}

\subsection{Main Ideas} \label{sec:ideas}

\paragraph*{Conditional Lower Bounds}
It is instructive to look at the conditional lower bounds. The reduction of 
triangle detection to subgraph counting in bounded degeneracy graphs is actually quite simple. Suppose we want to detect (or even count) triangles in an input graph $G$. Get
graph $G'$ by subdividing each edge into two, so a triangle in $G$
becomes a $\mcC_6$ ($6$-cycle) in $G'$. But the degeneracy of $G'$ is just 2!
(In any induced subgraph of $G'$, the minimum degree is at most $2$, proving
the bound.) Thus, if there exists $o(f(\degen)m^{1+\gamma})$ time algorithms
for counting $6$-cycles, that would violate the \TRICONJ. 

It is fairly straightforward to generalize this idea for larger cycles, by replacing
edges in $G$ by short paths. 
% Assuming \TRICONJ, for all $k \geq 6$,
% we can rule out linear time algorithms for counting $\mcC_k$ in bounded degeneracy graphs. We give the details in~\Cref{sec:hardness}.
Assuming \TRICONJ, for all $k \geq 6$ and $k\neq 8$,
we can rule out linear time algorithms for counting $\mcC_k$ in bounded degeneracy graphs. 
Our reduction does not work for $\mcC_8$; instead 
we consider a different subgraph for the case of $k=8$ ($\mcC_7$ with a tail).
We give the details in~\Cref{sec:hardness}.

This reduction fails for counting $5$-cycles and in general, it does not work
for counting any $5$-vertex subgraph. For good reason, as we discovered an
efficient algorithm for this problem. This is the more technical
part of our paper.

\paragraph*{Algorithmic Framework}
We present an algorithmic framework for solving the $\SUBC_k$ problem, that 
generalizes the core idea of Chiba and Nishizeki~\cite{chiba1985arboricity}.
It is known from past work that their ideas basically provide an $O(m\degen^{k-2})$ algorithm for 
$\SUBC_k$, for $k=3,4$. The main challenge is to get such an algorithm
for $k=5$, thereby nailing down the chasm of Theorem~\ref{thm:informal}.
This leads to new
results for counting various $5$-vertex subgraphs.  Perhaps more than these
individual results, our main contribution lies in identifying  structural 
decompositions of the pattern subgraphs that allows for efficient algorithms.
This decomposition also sheds light on why certain
$k$-vertex subgraphs, for $k\geq 6$, does not seem to have any
efficient algorithms in bounded arboricity graphs. We give an outline
of our framework next, and present it formally in~\Cref{sec:algorithm}.

The key idea that comes from Chiba-Nishizeki
is to perform subgraph counting on $G\dir$, an acyclic orientation of $G$ where the out degree of each vertex is bounded by $O(\degen)$\footnote{Technically, this is \emph{not}
the idea of Chiba-Nishizeki, who use the degree orientation. But it was somewhat
of a folklore result that it is easy to get the same result using the degeneracy
orientation. Arguably the first such reference is Schank-Wagener~\cite{ScWa05}.}.
The classic clique and 4-cycle counting algorithms enumerate directed stars
and directed paths of length $2$ to count subgraphs. We note that the algorithm
does \emph{not} enumerate 4-cycles, since there can be $\Omega(n^2)$ 4-cycles.
It requires clever indexing to solve this problem, which we generalize in our algorithm.

The crucial generalization of this idea is to enumerate directed rooted
trees. Specifically, we count occurrences of a connected pattern $H$ by counting occurrences of all possible acyclic orientations (up to isomorphism) $H\dir$ of $H$ in $G\dir$. The main idea is to find the largest directed rooted tree in $H\dir$, with edges directed away from the root. Call this tree $T$. Since outdegrees in $G\dir$ are bounded, we can efficiently enumerate all copies of $T$. Any copy of $H\dir$ in $G\dir$ is formed by extending a copy of $T$, but
$H\dir$ may contain vertices that are not in $T$. Thus, the extensions could
be expensive to compute.
But when $H$ has at most $5$ vertices, we can prove that $H\dir \setminus T$ 
is itself either a collection of rooted stars or paths. We can create
hash tables that store information about the occurrences of the latter. The final
count of $H\dir$ is obtained by enumerating $T$ and carefully combining counts
from the hash tables.

%%%%%%%%%%%%%%%%%%%%%%%%%%%%%%%%%%%%%%
%%%%%%%% RELATED WORK
%%%%%%%%%%%%%%%%%%%%%%%%%%%%%%%%%%%%%%
\section{Related Work}

Subgraph counting problems has a long and rich history.
More than three decades ago, Itai and Rodeh~\cite{itai1978finding} 
gave the first non-trivial algorithm for the triangle detection and 
counting problems with $O(m^{3/2})$ runtime. Subsequently, Chiba and 
Nishizeki~\cite{chiba1985arboricity} gave an elegant algorithm 
based on the degree based vertex ordering that solves
triangle counting, $4$-cycle counting and $\ell$-clique counting 
with running times of 
$O(m\degen)$, $O(m\degen)$, and $O(m\degen^{\ell -2})$ respectively 
($\degen$ denotes the degeneracy). In comparison, our algorithm exploits the 
\emph{degeneracy ordering} of the vertices (see~\Cref{sec:prelim} for a formal 
definition); this enables us to create a uniform framework for any
$k$-vertex subgraph for $k\in \{4,5\}$. 
% We remark that our framework can also 
% solve the $\ell$-clique counting problem, for any constant $\ell$, in
% $O(m\degen^{\ell-2})$ time. 
In dense graphs, the best bounds for the clique counting problem
are achieved by fast matrix multiplications based algorithms~\cite{nevsetvril1985complexity,eisenbrand2004complexity}; 
Vassilevska~\cite{vassilevska2009efficient} gave combinatorial algorithm
with significantly reduced space requirement. 
For general subgraphs, there is a rich line 
of research based on matrix multiplication, tree decomposition and
vertex cover methods~\cite{itai1978finding,nevsetvril1985complexity,alon1997finding,bjorklund2009counting,koutis2009limits,vassilevska2009finding,kowaluk2013counting,curticapean2014complexity,bjorklund2017counting,curticapean2017homomorphisms} --- these works focus on getting
$n^{\mu k}$ time algorithmis, for $\mu <1$.

Subgraph counting problems, specifically triangle counting, clique counting and 
cycle counting problems, has also been studied extensively in various Big Data models
such as property testing model~\cite{eden2017approximately,eden2018approximating,assadi2018simple}, 
MapReduce settings~\cite{cohen2009graph, Suri2011, kolda2014counting}, and streaming model~\cite{BarYossefKS02, Jowhari2005, Manjunath2011,Kane2012, ahn2012graph, Jha2013, Pavan2013, McGregor2016, bera2017towards}. 
Most of these work focuses on an approximate count, rather than an exact count. 
In the applied world, there are many efficient algorithms that are based on clever sampling techniques~\cite{BhRaRa+12,betzler2011parameterized,HoBe+07,zhao2012sahad,RaBhHa14,wernicke2006efficient,WernickeRasche06,jha2015path,wang2017moss}. Exact counting has also been studied extensively in the applied world~\cite{AhNe+15,PiSeVi17,ortmann2017efficient,birmele2012detecting,gonen2009approximating,milenkovic2008uncovering,stoica2009structure,marcus2010efficient,hovcevar2014combinatorial,hovcevar2016computation,hovcevar2017combinatorial,elenberg2015beyond,elenberg2016distributed}. In particular, Ahmed~\etal~\cite{AhNe+15} presented an algorithmic framework for solving the $\SUBC_4$ problem, called PGD (Parametrized Graphlet Decomposition), which scales to graphs with tens of millions of edges. Pinar~\etal~\cite{PiSeVi17} studied the $\SUBC_5$ problem, and gave the current state of the art ESCAPE library based on degree ordering techniques. However, the provable runtime of their algorithm for certain $5$-vertex subgraphs is quadratic, $O(n^2)$. For a deeper exploration of related applied work, refer to the tutorial on subgraph counting by Seshadhri and Tirthapura~\cite{SeTi19}.

The subgraph detection problem, which asks whether an input graph has a copy of the subgraph,
is a well-studied problem~\cite{itai1978finding,monien1985find,alon1995color,alon1997finding,kloks2000finding,kowaluk2013counting,williams2014finding}. 
For the triangle detection problem, the best known algorithm is based on 
fast matrix multiplication and it runs in time 
$O(\min \{n^\omega, m^{{2\omega}/{(\omega+1)}}\})$~\cite{alon1997finding}.
If $\omega=2$, this would give us $O(\min \{n^2, m^{4/3}\})$ algorithm for 
the triangle detection problem. Hence, to falsify the \TRICONJ, it would require a major
breakthrough result in the algorithmic graph theory world. For a
more detailed discussion on the \TRICONJ and its implications, refer to the paper by 
Abboud and Williams~\cite{abboud2014popular}.

In the subgraph enumeration problem, the goal is to output each occurrences of 
the target subgraph. Chiba and Nishizeki~\cite{chiba1985arboricity} showed that it is possible to
enumerate all the triangles in a graph along with counting the total number of triangles in $O(m\degen)$ time. For enumerating all the triangles, $O(m\degen)$ time is effectively optimal assuming the \threesum~\cite{patrascu2010towards,kopelowitz2016higher}.
Eppstein~\cite{eppstein1994arboricity} studied the bipartite 
subgraph enumaration problem in bounded arboricity graphs.

%%%%%%%%%%%%%%%%%%%%%%%%%%%%%%%%%%%%%%
%%%%%%%% PRELIM
%%%%%%%%%%%%%%%%%%%%%%%%%%%%%%%%%%%%%%
\section{Preliminaries}
\label{sec:prelim}

In this paper, we study the $\SUBC_k$ problem which
asks for the number of occurrences of each $k$-vertex subgraph $H$,
in an input graph $G$ with $n$ vertices and $m$ edges. We consider 
$k$ to be a constant.
For a fixed subgraph $H$, we use
$\SUBC_H$ to denote the problem of counting all occurrences of $H$ in the 
input graph $G$. When $H$ is the triangle subgraph, we denote the 
corresponding counting problem as $\TRIC$. In the context of the $\SUBC_k$ problem, we always use $G$ to denote the 
input graph and $H$ to denote the subgraph to be counted. 
Both $G$ and $H$ are simple, connected, undirected and unweighted. 

In our algorithmic framework, directed graphs play a crucial role. We use $N_G^+(u)$ and $N_G^-(u)$ to denote the out-neighborhood and in-neighborhood of a vertex $u$ in a directed graph $G$, respectively.
We define $d_G^+(u) = |N_G^+(u)|$ and $d_G^-(u) = |N_G^-(u)|$.
If the graph is clear from the context, we drop the subscript $G$. 

A graph $G$ is $k$-degenerate if every subgraph of $G$ has 
a vertex of degree at most $k$. The \emph{degeneracy} of a graph $G$
(also called coloring number, refer to Sec. 5.2 of~\cite{Diestel-book}),
denoted as $\degen(G)$, is the smallest
integer $k$ such that $G$ is $k$-degenerate. 
The \emph{arboricity} of a graph $G$, 
denoted as $\arbo(G)$, is the smallest integer $k$ such that the edge set $E(G)$
can be partitioned into $k$ forests. When the graph $G$ is clear from the context, 
we simply write $\degen$, and $\arbo$, instead of $\degen(G)$ and  $\arbo(G)$.
A classic theorem of Nash-Williams shows that 
the degeneracy and arboricity are closely related.
All our results can be stated in terms of
either of the parameters.
\begin{theorem} (Nash-Williams~\cite{Na64}) \label{fact:degen-arb}
  In every graph $G$,  $\arbo(G) \le \degen(G) \le 2\arbo(G)-1$. \qed
\end{theorem}

Vertex ordering is central to many subgraph counting algorithms. In
this paper, we work with the \emph{degeneracy ordering} of $G$, which is defined as follows.
\begin{definition} \label{def:degen-order}
\emph{Degeneracy ordering} of a graph $G$, denoted by $\before$, is obtained by repeatedly removing the vertex with minimum degree. The ordering is defined by the removal time. 
\end{definition}
\noindent For example, if $u \before v$, then $u$ is removed before $v$ according to the above process. 
\emph{Degeneracy ordering} can be found in linear time~\cite{matula1983smallest}. 

Using any vertex ordering $\prec$ of an undirected graph $G$, we construct a directed graph $G_\prec^\rightarrow$ as follows:
for each edge $\{u,v\} \in E(G)$, direct the edge from $u$ to $v$ iff $u \prec v$. We denote this directed edge as $(u,v)$.
Observe that $G_\prec^\rightarrow$ is necessarily acyclic. 
We denote the directed graph obtained from \emph{degeneracy ordering}
$\before$ as $G\dgnord$. The following two are folklore
results about vertex ordering and degeneracy, and can be derived
from Prop. 5.2.2 of~\cite{Diestel-book}.
\begin{lemma} \label{lem:degen-order-if}
For each vertex $v \in G\dgnord$, $d^+(v) \leq \degen$. \qed
\end{lemma}
\begin{lemma} \label{lem:degen-order-onlyif}
If there exists a vertex ordering $\prec$ of $G$ such that in the 
corresponding directed graph $G_{\prec}^{\rightarrow}$, $d^+(v) \leq k$ for each
vertex $v$, then $\degen(G) \leq k$. \qed
\end{lemma}
Next, we formally define a match (occurrence) of the target subgraph
$H$ in the input graph $G$. We also define a match in the context 
of directed graphs $H^\prime$ and $G^\prime$.
\begin{definition} \label{def:match}
A match of $H$ in $G$ is a bijection $\pi: S \rightarrow V(H)$ where $S \subseteq V(G)$ and for any two vertices $u$ and $v$ in $S$, $\{u, v\} \in E(G)$ if $\{\pi(u), \pi(v)\} \in E(H)$.
\end{definition}

\begin{definition} \label{def:dir-match}
A match of $H^\prime$ in $G^\prime$ is a bijection $\pi: S \rightarrow V(H^\prime)$ where $S \subseteq V(G^\prime)$ and for any ordered pair of vertices $(u,v)$ where $u$ and $v$ are in $S$, $(u, v) \in E(G^\prime)$ if $(\pi(u), \pi(v)) \in E(H^\prime)$.
\end{definition}

Our algorithm counts matches of $H$ in $G$ by counting matches of all possible acyclic orientations $H\dir$ of $H$ in $G\dgnord$. In general, whenever we use `$\dir$' to denote a directed graph, such as in $G\dgnord$ and $H\dir$, the directed graph is a DAG.

We denote the number of matches of $H$ in $G$ by $\matchCount(G,H)$. An incomplete match of $H$ in $G$ is an injection $\pi: S \rightarrow V(H)$ (so $|S| < |V(H)|$), that has the same properties of a match except being surjective.
% Consider a match $\pi$, if the mapping is not surjective, then we call $\pi$ an incomplete match. 
Consider two incomplete matches (injections) of $H$, $\pi_1: S_1 \rightarrow V(H)$, and $\pi_2: S_2 \rightarrow V(H)$. Let $V_{\pi_1} = \{\pi_1(u) \mid u \in S_1\}$ and $V_{\pi_2} = \{\pi_2(u) \mid u \in S_2\}$. We say that $\pi_2$ completes $\pi_1$ to be a match of $H$, when $V(H)=V_{\pi_1} \cup V_{\pi_2}$ (surjective), $V_{\pi_1} \cap V_{\pi_2} = \emptyset$ (injective), and for any two vertices $u \in S_1$ and $v \in S_2$, $\{u, v\} \in E(G)$ if $\{\pi_1(u), \pi_2(v)\} \in E(H)$. In case of directed graphs, it should hold that $(u, v) \in E(G^\prime)$ if $(\pi_1(u), \pi_2(v)) \in E(H^\prime)$ and $(v, u) \in E(G^\prime)$ if $(\pi_2(v), \pi_1(u)) \in E(H^\prime)$.

Two matches are distinct if they are not authomorphims of a match. In other words, two matches $\pi_1$ and $\pi_2$ of $H$ are equivalent, if they map two automorphisms of the exact same subgraph of $G$ to $H$. We denote the number of distinct matches of $H$ in $G$ by $\matchCountDist(G,H)$. In the $\SUBC_k$ problem, we are interested in $\matchCountDist(G,H)$ for all $k$-vertex subgraphs $H$.

%%%%%%%%%%%%%%%%%%%%%%%%%%%%%%%%%%%%%%
%%%%%%%% ALGORITHM
%%%%%%%%%%%%%%%%%%%%%%%%%%%%%%%%%%%%%%
\section{Subgraph Counting Through Orientation and Directed Trees}\label{sec:algorithm}
In this section, we discuss our algorithmic framework for solving the
$\SUBC_k$ problem. Instead of directly counting the number 
of occurrences of a $k$-vertex subgraph $H$ in the input graph
$G$, we count the occurrences of all possible DAG $H\dir$ (up to isomorphism) of $H$ in the graph $G\dgnord$.
To achieve this, our main idea 
is to find the largest directed tree of $H\dir$, enumerate all matches of this 
tree, and then count matches of the remaining vertices using structures we 
save in a hash table. 
In~\Cref{subsec:5vertex}, we show that our framework solves the
$\SUBC_5$ problem in expected $O(m\degen^3)$ time. In~\Cref{subsec:6vertex}, 
we demonstrate the limitation of our framework as it fails to solve 
the $\SUBC_{\mcC_6}$ problem \emph{efficiently}.

\begin{algorithm}[th]
\caption{Counting distinct matches of all 5-vertex subgraphs in $G$ ($\SUBC_5$)}\label{alg:counting-all}
\begin{algorithmic}[1]
\Procedure{Count-All-5}{$G$}
\State Derive $G\dgnord$ by orienting $E(G)$ with respect to degeneracy ordering.
\ForAll{connected 5-vertex subgraphs $H$ except 4-star}
% 4-star
% \If{$H$ is a 4-star}
% \State Save $\sum_{u \in V(G)} \binom{d(u)}{4}$
% \Else
% \State Run \textsc{Count-Match($G\dgnord, H$)} and save the result.
% \EndIf
\State Run \textsc{Count-Match($G\dgnord, H$)} and save the result for $H$.
\EndFor
\State Save $\sum_{u \in V(G)} \binom{d(u)}{4}$ for 4-star.
\EndProcedure
\end{algorithmic}
\end{algorithm}

\begin{algorithm}[th]
\caption{Counting distinct matches of $H$ in $G$ ($\SUBC_H$)}\label{alg:counting}
\begin{algorithmic}[1]
\Procedure{Count-Match}{$G\dgnord, H$}

% prelim
\State $\matchCountDist(G,H) \gets 0$
\ForAll{possible DAGs (up to isomorphism) $H\dir$ of $H$}
\State $\matchCount(G\dgnord,H\dir) \gets 0$
\State Find one of the largest $\drtss$s in $H\dir$, and call it $T_{\mm}$.

\ForAll{match $\pi$ of $T_{\mm}$ in $G\dgnord$}
\If{$\pi$ is a match of $H\dir$} \Comment{$V(T_{\mm})=V(H\dir)$. \Lem{match-check}}
\State $\matchCount(G\dgnord,H\dir) \gets \matchCount(G\dgnord,H\dir) + 1$
\ElsIf{$\pi$ is an incomplete match of $H\dir$} \Comment{\Lem{match-check}}
\State $k \gets$ number of ways to complete $\pi$ to a match of $H\dir$. \Comment{\Lem{match-completion}}
\State $\matchCount(G\dgnord,H\dir) \gets \matchCount(G\dgnord,H\dir) + k$
\EndIf
\EndFor
$\matchCountDist(G,H) \gets \matchCount(G\dgnord,H\dir)/|Aut(H\dir)|$
\EndFor
\State \Return $\matchCountDist(G,H)$
\EndProcedure
\end{algorithmic}
\end{algorithm}

\subsection{5-vertex Subgraph Counting}
\label{subsec:5vertex}
Our main algorithmic result is given in  the following theorem.

\begin{theorem}\label{thm:5-countability}
There is an algorithm that solves the $\SUBC_5$ problem in $O(m\degen^3)$ time.
\end{theorem}

Our strategy is to count matches of all possible DAGs (up to isomorphism) $H\dir$ of $H$ in $G\dgnord$, to obtain the number of distinct matches of $H$ in $G$. \Alg{counting} demonstrates this subroutine of our algorithm for $\SUBC_5$, which is shown in \Alg{counting-all}. First, we find one of the largest directed rooted tree subgraphs (\emph{$\drtss$}), which we define as follows, in $H\dir$.

\begin{definition}\label{def:drtss}
Given any directed graph $D$, a directed rooted tree subgraph ($\drtss$) of $D$, is a subgraph $T$ of $D$, where the underlying undirected graph of $T$ is a rooted tree, and edges are oriented away from the root in $T$.
\end{definition}

\noindent The following lemma shows that we can find all matches of any $\drtss$ in $H\dir$ in the desired time.

\begin{lemma}\label{lem:dir-tree-match}
Let $T$ be a directed tree with $k$ vertices. All matches of $T$ in $G\dgnord$ can be enumerated in $O(m\degen^{k-2})$.
\end{lemma}

\begin{proof}%[Proof of~\Cref{lem:dir-tree-match}]\label{proof:dir-tree-match}
Let $t_1,\ldots,t_k$ be a BFS ordering of $T$ starting at the root $t_1$.
%be the vertices of $T$, sorted by distance from its root, and then by ID (so, $t_1$ is root).
Fix an edge $(u,v) \in E(G\dgnord)$ and map $u$ to $t_1$ and $v$ to $t_2$. There are $m$ possible matches for $(t_1,t_2)$, which we can find by enumerating the edges of $G\dgnord$. Now, we will choose vertices to map to $t_3,\ldots,t_k$, one by one, in this order. Since the out-degree of each vertex in $G\dgnord$ is at most $\degen$, %$O(\arbo)$
if we have already mapped vertices to $t_1,\ldots,t_i$, there are at most $\degen$ %$O(\arbo)$
vertices that could be mapped to $t_{i+1}$. Therefor $\matchCount(G\dgnord,T) = O(m\degen^{k-2})$, and we can enumerate all of them by first choosing $(u,v)$ to map to $(t_1,t_2)$ and then choosing vertices to map to $t_3,\ldots,t_k$, in this order and one by one.
\end{proof}

\begin{observation}
Call a vertex $v$ of a directed graph a \emph{source} vertex, if $d^-(v)=0$. Consider $T$ to be one of the largest $\drtss$s of a DAG $D$. $T$ has to have a source vertex of $D$ as the root, otherwise the root has an in-neighbor $v$, which is not in $T$ as it would create a cycle. Adding $v$ to $T$ creates a new $\drtss$ which has one more vertex than $T$. This contradicts the fact that $T$ is one of the largest $\drtss$s of $D$. Hence, the root of $T$ has to be a source vertex of $D$.
\end{observation}

Given a 5-vertex DAG $H\dir$, we can find a $\drtss$ that has the most number of vertices among all $\drtss$s of $H\dir$ in constant time. First, find all source vertices, and then apply a Breath First Search (BFS) starting from each of these vertices and pick a BFS tree with the most number of vertices among all. The following lemma shows that the largest $\drtss$ has at least 3 vertices for a 5-vertex connected subgraphs, except 4-star. Notice that, the largest $\drtss$ of a 4-star with all the edges oriented towards the center has two vertices.

\begin{lemma}\label{lem:dir-tree-existence}
Let $H$ be a connected undirected 5-vertex graph that is not a 4-star. Each largest $\drtss$ of any DAG $H\dir$, which is an acyclic orientation of $H$, has at least three vertices.
\end{lemma}

\begin{proof}%[Proof of~\Cref{lem:dir-tree-existence}]\label{proof:dir-tree-existence}
We prove this lemma by contradiction. Assume that any $\drtss$ of $H\dir$ has at most two vertices. A directed 2-path, or any vertex with at least two outgoing edges result in a $\drtss$ with three vertices. Therefore,
\begin{enumerate}[(a)]
    \item $H\dir$ does not have a 2-path,
    \item each vertex in $H\dir$ has at most one outgoing edges.
\end{enumerate}
Notice that, since $H\dir$ is a DAG, it has at least one source vertex. Consider a source vertex $u$. Since $H$ is connected, $u$ has at least one neighbor, and by (b) it should have exactly one neighbor. Let $N^+(u)=\{v\}$, then $N^+(v)=\emptyset$, by (a). So, $v$ should have at least one incoming neighbor $w$. By (a), $w$ has no incoming edges, and it has no outgoing edges by (b). Call the other two vertices $x$ and $y$. As $H$ is connected, there should be a connection between $\{u,v,w\}$ and $\{x,y\}$. $u$ and $w$ cannot have any neighbor other than $v$, so $x$ and $y$ could only be connected to $v$. Since $H$ is not a star, there should be an edge between $x$ and $y$. Without loss of generality, let $(x,y)$ be that edge. By (a), $(y,v) \notin E(H\dir)$ and by (b) $(x,v) \notin E(H\dir)$. So, $\{u,v,w\}$ is not connected to $\{x,y\}$, and $H$ is disconnected, which is a contradiction. Thus, the assumption that any $\drtss$ of $H\dir$ has at most two vertices is wrong, and each largest $\drtss$ of $H\dir$ has at least three vertices.
\end{proof}

So far, we know that we can find one of the largest $\drtss$s of $H$, which has at least 3 vertices. We use $T_{\mm}$ to denote this $\drtss$. By \Lem{dir-tree-match}, we can enumerate all matches of $T_{\mm}$ in $G\dgnord$ in $O(m\degen^3)$ time. For each such match, we need to validate whether it is a (incomplete) match of $H\dir$ or not. If it is not, then it could not be completed to a match of $H\dir$. The following lemma shows that we can perform this validation efficiently. In the remaining part of this section, ``constanct expected time'', refers to constant amortized time access to hash maps that we use.

\begin{lemma}\label{lem:match-check}
Let $T$ be a $\drtss$ of a DAG $H\dir$ of a connected $k$-vertex graph $H$. Assume edges of $G\dgnord$ are saved in a hash table. For each match $\pi$ of $T$ in $G\dgnord$, it takes $O(|E(H\dir)|)$ expected time to validate whether $\pi$ is a (incomplete) match of $H\dir$ or not.
\end{lemma}

\begin{proof}%[Proof of~\Cref{lem:match-check}]\label{proof:match-check}
 Since $\pi$ is a bijection, it has an inverse which we denote by $\pi^{-1}$. Let $H\dir[V(T)]$ denote the subgraph of $H\dir$ induced on $V(T)$. Observe that, there could be edges in $H\dir[V(T)]$ not present in $T$. For $\pi$ to be a match (if $V(T)=V(H\dir)$) or incomplete match of $H\dir$, these edges have to be present between corresponding vertices in $G\dgnord$ mapped to $T$ by $\pi$. Formally, consider all ordered pairs of vertices $(a,b) \in V(T) \times V(T)$ such that $(a,b) \in E(H\dir)$ and $(a,b) \notin E(T)$, $\pi$ is a match or incomplete match of $H\dir$ iff  $(\pi^{-1}(a),\pi^{-1}(b)) \in E(G\dgnord)$ for all such pairs of vertices. To validate this, we enumerate all edges $(a,b)$ of $H\dir[V(T)]$ which are not present in $T$, and search for $(\pi^{-1}(a),\pi^{-1}(b))$ in hashed edges of $G\dgnord$ in expected constant time. So this only requires $O(|E(H\dir)|)$ expected time.
\end{proof}

If $V(T_{\mm})=V(H\dir)$, then a match of $T_{\mm}$ could be a match of $H\dir$ too, which could be verified as explained.  If there is a vertex in $H\dir$ which is not present in $T_{\mm}$, then after validating that a match of $T_{\mm}$ is an incomplete match of $H\dir$, we need to find the number of ways to complete it to a match of $H\dir$. For this we need to count matches of each possible structures that $T_{\mm}$ does not cover in $H\dir$. We save the count of these structures in $G\dgnord$, in hash tables. The following lemma shows that this can be done efficiently.

\begin{lemma}\label{lem:hashing}
In $O(m\degen^3)$ time and space, we can save all the following key and value pairs in hash maps $\hashmap_1$, $\hashmap_2$, and $\hashmap_3$.
\begin{enumerate}
\item $\hashmap_1: ((u,v),1)$ where $(u,v) \in E(G\dgnord)$
\item $\hashmap_2: (S, k)$ $\forall S \subseteq V(G\dgnord)$ where $1 \leq |S| \leq 4$, and $k$ is the number of vertices $u$ such that $S \subseteq N^+(u)$
\item $\hashmap_3: ((S_1,S_2),\ell)$ $\forall S_1, S_2 \subseteq V(G\dgnord)$, where $1 \leq |S_1 \cup S_2| \leq 3$, and $\ell$ is the number of edges $e=(u,v) \in E(G\dgnord)$ such that $S_1 \subseteq N^+(u)$ and $S_2 \subseteq N^+(v)$.
\end{enumerate}
\end{lemma}
\begin{proof}%[Proof of~\Cref{lem:hashing}]\label{proof:hashing}
We show how to enumerate and save all these structures in $\hashmap_1$, $\hashmap_2$, and $\hashmap_3$.
\begin{enumerate}
\item $\hashmap_1$: We can easily do this in $O(m)$ by enumerating the out-neighbors of each vertex 
\item $\hashmap_2$: For each edge $e=(u,v)$, we can enumerate all subsets $T$ of the set $\{w \in N^+(u) \mid v \before w\}$, % where $\before$ is the degeneracy ordering and
where $|T|\leq 3$, in $O(\degen^3)$ time,  and increment the value for the key $T \cup \{v\}$ in the hash map by one.
\item $\hashmap_3$: For each edge $e=(u,v)$ ($v \in N^+(u)$), we enumerate all possible subset $S_1 \subseteq N^+(u) \setminus \{v\}$ where $|S_1| \leq 3$. And, for each $S_1$ we enumerate all possible $S_2\setminus S_1$ in subsets of $N^+(v)$, such that $1 \leq |S_1 \cup S_2| \leq 3$. This takes $O(\degen^3)$ as the out-degree of each vertex is at most $\degen$, and we choose up to three vertices. All possible $S_1 \cap S_2$ can be determined by checking the connection between $v$ and each vertex in $S_1$ using the hashed edges of $G^\rightarrow$ in $\hashmap_1$. \qedhere
\end{enumerate} 
\end{proof}

The following lemma shows that we can count the number of ways to complete a match of $T_{\mm}$, which is also an incomplete match of $H\dir$, to a match of $H\dir$ efficiently.

\begin{lemma}\label{lem:match-completion}
Let $H$ be a 5-vertex connected graph, $H\dir$ be a DAG of $H$, and $T_{\mm}$ be one of the largest $\drtss$s in $H\dir$. Assume $\hashmap_1$, $\hashmap_2$, and $\hashmap_3$ are given. For each match $\pi$ of $T_{\mm}$ in $G\dgnord$ which is an incomplete match of $H\dir$, we can count the number of ways to complete $\pi$ to a match of $H\dir$ in expected constant time.
\end{lemma}

\begin{proof}%[Proof of~\Cref{lem:match-completion}]\label{proof:match-completion}
By \Lem{dir-tree-existence}, $T_{\mm}$ has at least 3 vertices, and since $\pi$ is an incomplete match (not a match) of $H\dir$, we can assume that $|V(T_{\mm})| < 5$. Observe that, $T_{\mm}$ is a maximal $\drtss$. Any vertex in $H\dir$ which is not in $T_{\mm}$ can only be connected to vertices of $T_{\mm}$ by outgoing edges, otherwise they could be added to $T_{\mm}$ to create a larger $\drtss$ of $H\dir$, which contradicts the maximality of $T_{\mm}$. We consider two cases where $T_{\mm}$ has three or four vertices.

Let $|V(T_{\mm})| = 4$, and $i$ be the only vertex in $H\dir$ that is not in $T_{\mm}$. To complete $\pi$ to a match of $H\dir$, we need to choose a vertex in $G\dgnord$, that is connected by outgoing edges to vertices mapped to the out-neighborhood of $i$ in $H\dir$. Let $S_i = \{\pi^{-1}(t) \mid t \in N_{H\dir}^+(i)\}$. $\hashmap_2(S_i)$ is the number of vertices that could be mapped to $i$, but some of them may be already mapped to a vertex in $T_{\mm}$, by $\pi$. Let $r_i$ denote the number of vertices $v \in \{\pi^{-1}(t) \mid t \in V(T_{\mm})\}$, where $S_i \subseteq N_{G\dgnord}^+(v)$. We can obtain $r_i$ in expected constant time, by enumerating vertices mapped to $V(T_{\mm})$, and counting vertices that are connected to all vertices in $S_i$. For any vertex, we can check the connection to each vertex of $S_i$ using $\hashmap_1$ in expected constant time.
The number of ways to complete $\pi$ to a match of $H\dir$ in this case is $\hashmap_2(S_i)-r_i$.

Now we consider the case where $|V(T_{\mm})| = 3$. Let $V(H\dir)\setminus V(T_{\mm}) = \{i,j\}$. To complete $\pi$ to a match of $H\dir$, we only need to choose two vertices of $G\dgnord$ to map to $i$ and $j$. Let $S_i = \{\pi^{-1}(t) \mid t \in V(T_{\mm}) \cap N_{H\dir}^+(i)\}$ and $S_j = \{\pi^{-1}(t) \mid t \in V(T_{\mm}) \cap N_{H\dir}^+(j)\}$. We consider two cases, where $i$ and $j$ are connected or not. If they are connected, without loss of generality, assume $(i,j) \in E(H\dir)$. If $(i,j) \in E(H\dir)$, then we can use $\hashmap_3$ in \Lem{hashing}, to find the number of edges $(u,v)$ where $u$ and $v$ could be mapped to $i$ and $j$, respectively. Let $r_{(i,j)}$ be the number of edges $e=(w,x)\in E(G\dgnord)$, where $w$ and $x$ are mapped to vertices in $T_{\mm}$ by $\pi$, such that, $S_i \subseteq N_{G\dgnord}^+(w)$, and $S_j \subseteq N_{G\dgnord}^+(x)$. We can obtain $r_{(i,j)}$ in expected constant time using $\hashmap_1$. Then the number of edges $(u,v)$ that could be mapped to $(i,j)$ is $\hashmap_3((S_i, S_j))-r_{(i,j)}$. Next case is when $(i,j) \notin E(H\dir)$. In this case, we use $\hashmap_2$ to find the number of pair of vertices of $G\dgnord$ which could be mapped to $i$ and $j$. Let $r_i$ ($r_j$ resp.) denote the number of vertices $v \in V(G\dgnord)$ where $v$ is mapped to a vertex in $T_{\mm}$ and $S_i \subseteq N_{G\dgnord}^+(v)$ ($S_j \subseteq N_{G\dgnord}^+(v)$ resp.). Also, we use $r_{i,j}$ to denote the number of vertices $v \in V(G\dgnord)$ that are counted in both $r_i$ and $r_j$, meaning $S_i \cup S_j \subseteq N_{G\dgnord}^+(v)$. We can obtain $r_i$, $r_j$, and $r_{i,j}$ easily in expected constant time using $\hashmap_1$. The number of pairs of vertices which could be mapped to $i$ and $j$ is equal to $(\hashmap_2(S_i) - r_i) \cdot (\hashmap_2(S_j)-r_j) - (\hashmap_2(S_i \cup S_j)-r_{i,j} )$.
\end{proof}

Now, we have all the tools to efficiently count distinct matches of a DAG of $H\dir$ in $G\dgnord$. The following lemma shows that we can do this in $O(m\degen^3)$ expected time.

\begin{lemma}\label{lem:5-dag-countability}
There is an algorithm which counts distinct matches for each possible DAG (up to isomorphism) $H\dir$ of a 5-vertex connected subgraphs $H$, in $O(m\degen^3)$ expected time.
\end{lemma}

\begin{proof}%[Proof of~\Cref{lem:5-dag-countability}]\label{proof:5-dag-countability}
Fix a DAG $H\dir$ of $H$. If $H$ is a 4-star and $H\dir$ has $\ell$ incoming neighbors, then the number of distinct matches of $H\dir$ is $\sum_{u \in V(G\dgnord)} \binom{d^-(u)}{\ell} \binom{d^+(u)}{4-\ell}$. Assume that $H$ is not a 4-star. Find a $\drtss$ of $H\dir$ with the most number of vertices among all its $\drtss$s, and call it $T_{\mm}$. This can be done in constant time for $H\dir$. By \Lem{dir-tree-existence}, $T_{\mm}$ has at least three vertices. We will now enumerate all matches of $T_{\mm}$ in $G\dgnord$. By \Lem{dir-tree-match}, this step requires $O(m\degen^3)$ expected time. For each match $\pi$ of $T_{\mm}$ in $G\dgnord$, we can verify whether $\pi$ is a match (if ($|V(T_{\mm})|=5$) or incomplete match of $H\dir$ in expected constant time, by \Lem{match-check}. If $|V(T_{\mm})|=5$, while enumerating all matches of $T_{\mm}$, we only count them if they are a match of $H\dir$. So in this case we can count $\matchCount(G\dgnord, H\dir)$ in $O(m\degen^3)$ expected time. 

Otherwise, $T_{\mm}$ has 3 or 4 vertices. In this case, for each match $\pi$ of $T_{\mm}$, we first verify that it is also an incomplete match of $H\dir$. Then, we count the number of ways to complete $\pi$ to a match of $H\dir$, which we can do in expected constant time, by \Lem{match-completion}. To obtain $\matchCount(G\dgnord, H\dir)$, we simply sum the ways to complete each incomplete match we have found, to a match of $H\dir$.

This approach gives us the number of all (not necessarily distinct) matches of $H\dir$ in $G\dgnord$. Let $H\dir_\pi$ be a subgraph of $G\dgnord$ that $\pi$ maps to $H\dir$. Each automorphism of $H\dir$, gives a new match $\pi^\prime$ which is not distinct from $\pi$, as it is still mapping $H_\pi$ (the same copy of $H$) to $H\dir$ (example in \Fig{5-vertex-example-equiv-matches}). As each match of $H\dir$, also maps vertices to $T_{\mm}$, resulting in a match of $T_{\mm}$ and an (incomplete) match of $H\dir$, we will find all distinct matches of $H\dir$ and count each one exactly $|Aut(H\dir)|$ times. We want the number of distinct matches, which we can obtain by dividing the count of all matches by $|Aut(H\dir)|$.

Thus, it requires $O(m\degen^3)$ expected time to create $\hashmap_1$, $\hashmap_2$, and $\hashmap_3$ by \Lem{hashing}, $O(m\degen^3)$ time for enumerating matches of $T_{\mm}$, expected constant time to validate these matches, and expected constant time for counting ways to complete each such match, that is verified to be an incomplete match of $H\dir$, to a match of $H\dir$. So overall, we can find $\matchCountDist(G\dgnord, H\dir)$ in $O(m\degen^3) $ expected time.

This completes the proof of this lemma.
\end{proof}

\begin{figure*}[t]
\begin{subfigure}[t]{\textwidth}
\centering
  \begin{tikzpicture}[nd/.style={scale=1,circle,draw,fill=graph-vertex-blue,inner sep=2pt},minimum size = 6pt]
    \matrix[column sep=1cm, row sep=0.2cm,ampersand replacement=\&]
    {
     \FiveVertexExample\&
     \DirFiveVertexExample\\
    };
  \end{tikzpicture}
\caption{$H$ is 5-vertex connected subgraph and $H\dir$ is one possible acyclic orientation of it. $T_{\mm}$ (largest $\drtss$ of $H\dir$) is shown in green and contains three vertices.}
\label{fig:5-vertex-example-dag}
\end{subfigure}\\
\begin{subfigure}[t]{\textwidth}
\resizebox{\textwidth}{!}{
  \begin{tikzpicture}[nd/.style={scale=1,circle,draw,fill=graph-vertex-blue,inner sep=2pt},minimum size = 6pt]
    \matrix[column sep=0.8cm, row sep=0.5cm,ampersand replacement=\&]
    {
     \DirFiveVertexExampleOneMatch \&
     \DirFiveVertexExampleTwoMatch \&
     \DirFiveVertexExampleThreeMatch \\
     \DirFiveVertexExampleFourMatch \&
     \DirFiveVertexExampleFiveMatch \&
      \DirFiveVertexExampleSixMatch \\
    };
  \end{tikzpicture}
  }
\caption{All six figures show exactly the same subgraph in $G\dgnord$. $\pi_1,\ldots,\pi_6$ are six equivalent matches of $H\dir$ in $G\dgnord$, one for each automorphism of $H\dir$. Notice $(u,v,w)$ being mapped to all permutations of $(a,b,c)$.}
\label{fig:5-vertex-example-equiv-matches}
\end{subfigure}
\caption{Application of \Alg{counting} on a DAG $H\dir$ of an example 5-vertex connected subgraph $H$.}
\label{fig:5-vertex-example}
\end{figure*}

Lastly, we can prove \Thm{5-countability} as follows.

\begin{proof}[Proof of~\Cref{thm:5-countability}]\label{proof:5-countability}
Given a 5-vertex connected subgraph $H$, we can count all distinct matches of each possible DAG $H\dir$ of $H$, in $G\dgnord$ in $O(m\degen^3)$ expected time, by \Lem{5-dag-countability}. To count all distinct matches of $H$ in $G$, we just need to sum the number of distinct matches of all possible DAGs (up to isomorphism) of $H$. The number of such DAGs is constant for $H$. There are 21 different connected 5-vertex subgraphs (illustrated in~\cite{PiSeVi17}), and we perform this process on all of them.
This completes the proof of the theorem.
\end{proof}

\subsection{Limitations of Our Framework for a Six Vertex Subgraph}
\label{subsec:6vertex}

\begin{figure*}[t]
\centering
\begin{tikzpicture}[nd/.style={scale=1,circle,draw,fill=blue,inner sep=2pt},minimum size = 6pt]    
    \matrix[column sep=1cm, row sep=0.2cm,ampersand replacement=\&]
    {
     \UncountableSixVertexExample\\
 };
  \end{tikzpicture}
\caption{Let $H\dir$ be a DAG of $H$ ($\mcC_6$). Considering any largest $\drtss$ of $H\dir$, the remaining vertices include a vertex with two incoming edges (in-in wedge). Even graphs with bounded degeneracy can have $\Omega(n^2)$ in-in wedges. So hashing in \Alg{counting} will not be bounded by $m$ and $\degen$ for $H$.}
\label{fig:6-vertex-uncountable}
\end{figure*}
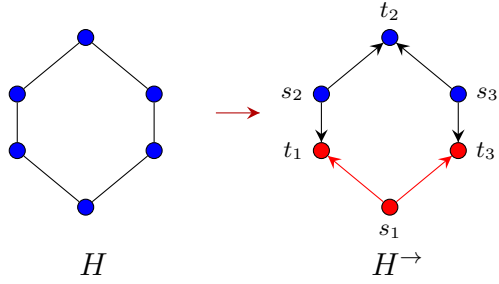

Consider $\mcC_6$, shown as $H$ in \Fig{6-vertex-uncountable}. Then $H\dir$, shown in the right side of \Fig{6-vertex-uncountable}, is a possible DAG of $H$. In $H\dir$, $s_1$, $s_2$, and $s_3$ are the source vertices, and $t_1$, $t_2$, and $t_3$ are the sink vertices. Any $\drtss$ of $H\dir$ has at most three vertices, and there are three such $\drtss$, $T_1$, $T_2$, and $T_3$ rooted at $s_1$, $s_2$ and $s_3$, respectively. $T_1$ is shown by red in \Fig{6-vertex-uncountable}. For each of $T_1$, $T_2$, and $T_3$, the remaining vertices include a vertex, with two incoming edges, which we call an in-in wedge. For example, $t_2$ is such a vertex for $T_1$. Even graphs with bounded degeneracy can have $\Omega(n^2)$ in-in wedges.  
We cannot hash the count of such structures in expected time bounded by $m$ and $\degen$. So, \Alg{counting} fails to count occurrences of $\mcC_6$ in the desired time. In the next 
section, we discuss why such limitations are natural to any framework for the $\SUBC_k$ problem at and beyond $k=6$.

%%%%%%%%%%%%%%%%%%%%%%%%%%%%%%%%%%%%%%
%%%%%%%% HARDNESS
%%%%%%%%%%%%%%%%%%%%%%%%%%%%%%%%%%%%%%
\section{A Chasm at Six} \label{sec:hardness}
At the end of the previous section, we showed the limitations
of our framework in counting certain $6$-vertex subgraphs. In 
this section, we show that perhaps such limitations are
fundamental to any subgraph counting algorithms.
In particular, the landscape  of
$\SUBC_k$ problem in the bounded degeneracy graphs changes 
dramatically as we move beyond $k=5$.
We prove that for every integer $k\geq 6$,
there exists a $k$-vertex subgraph $H$ such that, the 
running time of any algorithm for the $\SUBC_H$ problem
does not depend on the degeneracy of the input graph, assuming
the \TRICONJ. In contrast, for $k\leq 5$,
$O(m\degen^{k-2})$ algorithms exists for $\SUBC_k$ (see ~\Cref{sec:algorithm}). The following theorem captures the main result
of this section.
\begin{theorem}
\label{thm:conditional_LB}
Assume the \TRICONJ (Conj. \ref{conj:triangle}). There exists an absolute constant $\gamma > 0$ such that the following
holds. For any $k \geq 6$ and any function $f:\mathbb{N} \to \mathbb{N}$, there exists a $k$-vertex subgraph $H$ such that
there is no (expected) $o(m^{1+\gamma}f(\degen))$ algorithm for $\SUBC_H$.
\end{theorem}

\subparagraph*{Outline of the Proof}
%In the remaining of this section, we prove~\Cref{thm:conditional_LB}.
For each $k\geq 6$ and $k\neq 8$, the subgraph of interest will be the $k$-cycle graph, 
$\mcC_k$. For $k=8$, the subgraph of interest will be the $\mcC_7$ with a tail (see~\Cref{fig:tailed-C7}).
We first give a proof outline. Fix some $k\geq 6$ and
let $H_k$ denote the target subgraph of size $k$. 
Recall the $\TRIC$ problem ---
count the number of triangles in a graph with $m$ edges. \Cref{conj:triangle}
asserts that for any algorithm $\mcA$ for the $\TRIC$ problem, $T(\mcA) = 
\omega(m)$ where $T(\mcA)$ denotes the worst case time complexity of 
the algorithm $\mcA$.
Our strategy is to reduce from the $\TRIC$ problem to the
$\SUBC_{H_k}$ problem. To this end, we construct
a new graph $G_k$ from the input instance $G$ of the $\TRIC$ problem
such that $G_k$ has $O(m)$ edges, and 
has degeneracy at most $2$. More importantly,
the number of triangles in $G$ is a simple linear function of the number
of $H_k$ in $G_k$.
Hence, we can derive the number of triangles in $G$ by counting
the number of $H_k$ in $G_k$.
As $\degen(G_k) \leq 2$, any $O(m f(\degen))$ algorithm for the $\SUBC_{H_k}$ problem translates to a $O(m)$ algorithm for the $\TRIC$
problem, contradicting the \TRICONJ. 
We remark that, for $k=8$, our proof strategy will
be slightly different --- instead of 
reducing from the $\TRIC$ problem, we shall reduce from
the triangle detection problem itself. However,
the gadget construction will follow the same basic
principle. 

The construction of $G_k$ from $G$ is rather simple. 
The details of the construction depends on whether $k$
is a multiple of $3$ or not. 
We take two examples to describe the construction. 

First, we take $k=6$, and the target subgraph $H_6=\mcC_6$.
For each edge $e$ in $E(G)$, we replace $e$ with a length two path 
$\{e_1,e_2\}$ in $E(G_6)$. 
To accomplish this, we add a new vertex $v_e$
for each edge: $V(G_6) = V(G) \cup \{v_e\}_{e\in E(G)}$.
This is shown in~\Cref{fig:edge-split-C6}.
Each triangle in $G$ creates a $\mcC_6$
in $G_6$. We formally prove in~\Cref{lem:count_func}
that the number of triangles in $G$ is
same as the number of $\mcC_6$ in $G_6$. In~\Cref{lem:degen},
we bound the degeneracy of $G_6$ by $2$.
This construction can be generalized for any $k=3\ell$ where 
$\ell \geq 2$, by replacing each edge in $E(G)$ with $\ell$-length 
path.

Next consider the case $k=7$. 
% The same construction works for $k=8$ as well. 
For each edge $e\in E(G)$, we first create two 
parallel copies of $e$, and then replace the first one with 
a length two path $\{e_{1,1}$, $e_{1,2}\}$, and the second one with a 
length three path $\{e_{2,1}$, $e_{2,2}$, $e_{2,3}\}$. 
So in $E(G_7)$, we have $5$ edges for each edge in $E(G)$. 
We create $3$ new vertices per edge
to accomplish this, and denote them as $v_e,u_{e_1},u_{e_2}$. See~\Cref{fig:edge-split-C7} for a pictorial demonstration. In~\Cref{lem:count_func}, we argue that the number of 
$\mcC_7$ is exactly $3$ times the number of 
triangles in $G$. In~\Cref{lem:degen},
we bound the degeneracy of $G_7$ by $2$. This construction generalizes to any $k=3\ell+i$
where $\ell \geq 2$ and $i \in \{1,2\}$ (except for the case when $k=8$, that is $\ell =2$ and $i=2$) by splitting each edge 
into $\ell$ and $\ell+1$ many parts respectively.

Finally, we consider the case of $k=8$. Note that the target subgraph
$H_8$ is the $7$-cycle with a tail in this case (see ~\Cref{fig:tailed-C7}).
It is natural to wonder why do we not simply take $H_8=\mcC_8$?
After all, for all other values of $k$, taking $H_k=\mcC_k$ suffices.
At a first glance, it seems like if we consider the same graph $G_7$
as described above (and in~\Cref{fig:edge-split-C7})
the number of $\mcC_8$ would be a
simple linear function of the number of triangles in $G$
--- for each triangle in $G$, there will be exactly three $\mcC_8$ in
$G_7$. However, each $\mcC_4$ in $G$ would also lead to a $\mcC_8$ in $G_8$.
Observe that for $k > 8$, we do not run into this problem. A more formal
treatment of this issue appear in~\Cref{proof:LB}.

So instead, we take $H_8$ to be the subgraph $\mcC_7$ with a tail to prove
our conditional lower bound for $\SUBC_8$. The construction of
the graph $G_8$ remains exactly the same as that of $G_7$. We show in~\Cref{lem:cycle_with_tail} that, there exists a $\mcC_7$ with a tail
in $G_8$ if and only if there exist a triangle in $G$. 

\begin{figure*}[h]
    \centering
    \begin{tikzpicture}[nd/.style={scale=1,circle,draw,fill=blue,inner sep=2pt},minimum size = 6pt]    
      \foreach \l [count=\n] in {0,1,2,3,4,5,6} {
        \pgfmathsetmacro\angle{90-360/7*(\n-1)}
        \ifnum\n=2
            \node[nd] (a\n) at (\angle:1.5) {};
            % \node[nd] (b\n) at (\angle:2.25) {};
        \else
            \node[nd] (a\n) at (\angle:1.5) {};
        \fi
          
      }
      \draw (a6) -- (a7) -- (a1) -- (a2) -- (a3) -- (a4)-- (a5)-- (a6);
    %   \draw (a2) -- (b2);
       \pgfmathsetmacro\angle{-336}
       \node[nd] (c) at (\angle:2.3) {};
       \draw (a2) -- (c);
    \end{tikzpicture}
    \caption{Target subgraph for proving conditional lower bounds for $\SUBC_8$: the $\mcC_7$ with a tail}
    \label{fig:tailed-C7}
\end{figure*}
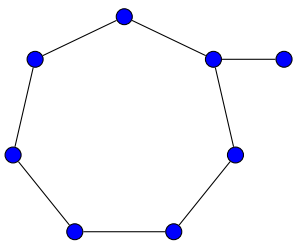

\begin{figure*}[t]
\begin{subfigure}[t]{0.47\textwidth}
\centering
\resizebox{\textwidth}{!}{
\begin{tikzpicture}[nd/.style={scale=1,circle,draw,fill=blue,inner sep=2pt},minimum size = 6pt]    
    \matrix[column sep=1cm, row sep=0.2cm,ampersand replacement=\&]
    {
     \EdgeSplitSixCycle \\
 };
  \end{tikzpicture}
}
\caption{Construction of the edge set $E(G_6)$ from the edge set $E(G)$. The red colored nodes are only present in $V(G_6)$, and not in $V(G)$.}
\label{fig:edge-split-C6}
\end{subfigure}\hfill
\begin{subfigure}[t]{0.47\textwidth}
\centering
\resizebox{\textwidth}{!}{
\begin{tikzpicture}[nd/.style={scale=1,circle,draw,fill=blue,inner sep=2pt},minimum size = 6pt]    
    \matrix[column sep=1cm, row sep=0.2cm,ampersand replacement=\&]
    {
     \EdgeSplitSevenCycle \\
 };
  \end{tikzpicture}
}
\caption{Construction of the edge set $E(G_7)$ from the edge set $E(G)$. The red colored nodes are only present in $V(G_7)$, and not in $V(G)$.}
\label{fig:edge-split-C7}
\end{subfigure}
\caption{Reduction from the \TRIC problem to the $\SUBC_{\mcC_k}$ problem for $k=6$ (left) and $k=7$ (right).}
\label{fig:reduction}
\end{figure*}

We now present the proof of ~\Cref{thm:conditional_LB} in full details.

\begin{proof}[Proof of~\Cref{thm:conditional_LB}]\label{proof:LB}
Fix some $k\geq 6$. Let the subgraph $H_k$ denote the target subgraph of size $k$. For $k\neq 8$, $H_k$
is $\mcC_k$, and for $k=8$, $H_k
$ is $\mcC_7$ with a tail (see~\Cref{fig:tailed-C7}).
We reduce from the \TRIC problem to the $\SUBC_{H_{k}}$.
Let $G=(V,E)$ be the input instance for the \TRIC problem 
with $|V|=n$ and $|E|=m$. We construct an input instance $G_{k}=(V_{k},E_{k})$
for the $\SUBC_{H_{k}}$ problem from $G$.
The construction of $G_k$ differs based on whether 
$k$ is divisible by $3$ or not. We next consider these two cases separately.

\subparagraph*{Details of the Reduction.} First assume $k=3\ell$ for some integer $\ell \geq 2$. 
We first define the vertex set $V_k$. 
For each vertex in $V$, we add a vertex in $V_k$. 
For each edge $e \in E$, we add a set of $\ell -1$ many vertices,
denoted as $V_e = \{v_{e,1},v_{e,2},\ldots,v_{e,\ell-1}\}$.
We collect all these second type of vertices into the set $V_E$.
Formally, we have
\begin{align*}
    V_k &= V \cup V_E \,, \\
    \text{where } V_E &= \bigcup_{e \in E} V_e \,, \\
    \text{for } V_e &= \{v_{e,1},v_{e,2},\ldots,v_{e,\ell-1}\} \,.
\end{align*}
We now describe the edge set $E_k$. We treat 
each edge $e=\{u,v\} \in E$ as an ordered pair $(u,v)$
where the ordering can be arbitrary of the vertices (for example, assume
lexicographical ordering). Now for each edge $e=(u,v)$
construct an $\ell$-length path between $u$ and $v$ in $V_k$
by connecting the vertices in $\{u\}\cup V_e\cup \{v\}$ sequentially. 
More precisely, we define $E_k$ as follows.
\begin{align*}
E_k &= \bigcup_{e \in E} E_e \,, \\
\text{where } E_e &= \{ \{u,v_{e,1}\}, \{v_{e,1},v_{e,2}\}, \ldots, 
\{v_{e,\ell-2},v_{e,\ell-1}\},\{v_{e,\ell-1},v\} \}\text{ for $e=(u,v)$} \,.
\end{align*}
This completes the construction of the graph $G_k=(V_k,E_k)$. 
We give an example in ~\Cref{fig:edge-split-C6} for $k=6$.

Now assume $k=3\ell + i$ for some some integer $\ell \geq 2$ and $i\in \{1,2\}$.
In the previous case, we added a set of $\ell -1$ many vertices for each 
edge in $E$.
But now, for each edge $e\in E$, we add two sets of vertices, one with $\ell-1$
many vertices and the other with $\ell$ many vertices. We denote 
the first set as $V_e = \{v_{e,1},v_{e,2},\ldots,v_{e,\ell-1}\}$, and
the second set as $U_e = \{u_{e,1},u_{e,2},\ldots,u_{e,\ell}\}$.
We also add the set of vertices in $V$ to $V_k$.
Formally, we have 
\begin{align*}
    V_k &= V \cup V_E \,, \\
    \text{where } V_E &= \bigcup_{e \in E} V_e \cup U_e\,, \\
    \text{for } V_e &= \{v_{e,1},v_{e,2},\ldots,v_{e,\ell-1}\} \,, \\
    \text{and } U_e &= \{u_{e,1},u_{e,2},\ldots,u_{e,\ell}\} \,.
\end{align*}
To construct the edge set $E_k$, as before we treat each edge 
in $e=\{u,v\} \in E$ as an ordered pair $(u,v)$ according to some
arbitrary ordering of the vertices. Now, for each edge $e=(u,v)$, construct an $2\ell+1$-length cycle between $u$ and $v$ in $V_k$
by creating a $\ell$-length path via the vertices in $V_e$ 
and another $\ell+1$-length path via the vertices in $U_e$.
We denote the corresponding edge sets as $E_{V,e}$ and 
$E_{U,e}$ respectively. Formally, we define $E_k$ as follows.
\begin{align*}
E_k &= \bigcup_{e \in E} \left(E_{V,e} \cup E_{U,e} \right) \,, \\
\text{where } E_{V,e} &= \{ \{u,v_{e,1}\}, \{v_{e,1},v_{e,2}\}, \ldots, 
\{v_{e,\ell-2},v_{e,\ell-1}\},\{v_{e,\ell-1},v\} \} \,, \\
\text{and } E_{U,e} &= \{ \{u,u_{e,1}\}, \{u_{e,1},u_{e,2}\}, \ldots, 
\{u_{e,\ell-1},u_{e,\ell}\},\{u_{e,\ell},v\} \} \text{ for $e=(u,v)$} \,.
\end{align*}
This completes the construction
of the graph $G_k=(V_k,E_k)$. Note that the construction 
is independent of the value of $i$. Hence, we produce 
the same graph $G_k$ for $k=3\ell+1$ and $k=3\ell+2$. 
We give an example in~\Cref{fig:edge-split-C7} for $k=7$.

Note that although our target subgraph for the case $k=8$ is
a $7$-cycle with a tail instead of $8$-cycle, our construction is still
the same.

\subparagraph*{Correctness of the Reduction}
In~\Cref{lem:degen}, we prove that $G_k$ has degeneracy at most $2$.
In~\Cref{lem:count_func}, we show that, for $k\neq 8$, the number of $\mcC_k$ in 
the graph $G_k$ is a linear function of the number of triangles in $G$. 
In~\Cref{lem:cycle_with_tail}, we show that $G_8$ is $H_8$ free if and only
if $G$ is triangle free.
\begin{lemma}\label{lem:degen}
$\degen(G_k) \leq 2$.
\end{lemma}
\begin{proof}
To prove the lemma it is sufficient to exhibit a vertex ordering 
$\prec$ such that in the corresponding directed graph 
$G_{\prec}^{\rightarrow}$, $d^{+}(v) \le 2$ for all $v \in V_k$ (application of~\Cref{lem:degen-order-onlyif}).
We use an ordering $\prec$ where $V_E \prec V$ and the ordering within each
set is arbitrary. Observe that each vertex $v \in V_E$ has degree exactly $2$ and
no two vertices in $V$ are connected to each other. Hence,
$d^{+}(v) \le 2$ for all $v \in V_k$.
\end{proof}
\begin{lemma}\label{lem:count_func}
Let $\ell \geq 2$ be some integer.
For $k=3\ell$, $\matchCountDist(G_k,\mcC_k)= \matchCountDist(G,\mcC_3)$. 
For $k=3\ell+i$ with $i\in \{1,2\}$ and $k\neq 8$, 
$\matchCountDist(G_k,\mcC_k)= 3\cdot  \matchCountDist(G,\mcC_3)$. 
\end{lemma}
\begin{proof}
Let $\mcT$ be the set of triangles in $G$ and $\mcC$ be the set of
$\mcC_k$ in $G_k$. Note that a triangle in $\mcT$ and a $k$-cycle in $\mcC$
can be uniquely identified by a set of three and $k$ edges, respectively.

We first take up case of $k=3\ell$ for some $\ell \geq 2$. Let $g$ be the mapping between the sets $\mcT$ and $\mcC$, $g:\mcT \rightarrow \mcC$, defined as follows:
$g(\{e_1,e_2,e_3\}) = E_{e_1} \cup E_{e_2} \cup E_{e_3}$. To prove the lemma,
it is sufficient to exhibit that $g$ is a bijection. To this end, note 
that if $g(\tau_1) = g( \tau_2)$, then $\tau_1 = \tau_2$. 
This follows immediately from the definition of $g$, 
since $E_{e_1} \cap E_{e_2} = \emptyset$ for all $e_1\neq e_2$. We now
show that every $k$-cycle in $\mcC$ has an inverse mapping in $g$.
Let $\xi$ be a $k$-cycle in $\mcC$. Fix some edge $e\in E$. By construction,
either all the edges from the set $E_e$ are present in $\xi$, or none
of them are. Hence, $\xi$ must be
of the form $E_{e_1} \cup E_{e_2} \cup E_{e_3}$ for some
three distinct edges $e_1$, $e_2$, and $e_3$. Clearly, $\{e_1,e_2,e_3\}$
forms a triangle in $G$.

Now assume $k=3\ell+i$ for some $\ell\ge 2$ and $i\in \{1,2\}$, and $k\neq 8$.
It is not difficult to see that each triangle in $\mcT$ 
leads to exactly three $k$-cycles in $\mcC$. The non-trivial
direction is to show that for each $k$-cycles in $\mcC$
there is an unique triangle in $\mcT$. 
Let $\xi$ be a $k$-cycle in $\mcC$. 
Fix some edge $e\in E$. By construction,
exactly one of the following must be true:
(i) all the $\ell$ edges from the set $E_{V,e}$ are present in $\xi$,
(ii) all the $\ell+1$ edges from the set $E_{U,e}$ are present in $\xi$,
(iii) none of the edges from the set $E_{V,e} \cup E_{U,e}$ 
are present in $\xi$. 
First assume $i=1$. Since $\xi$ has $3\ell+1$ many edges,
and $\ell \geq 2$, it must consist of one $E_{U,e}$ set of size $\ell+1$, and two
$E_{V,e}$ sets of size $\ell$.  When $i=2$ and $\ell > 2$,
$\xi$ must consist of two $E_{U,e}$ set of size $\ell+1$, and one
$E_{V,e}$ sets of size $\ell$. 
Clearly, the three edges corresponding to
these sets form a unique triangle in $G$.
(When $k=8$, that is $\ell=2$ and $i=2$, taking four distinct sets $E_{V,e}$ 
creates a copy of $\mcC_8$, and hence the argument does not work.)
\end{proof}
\begin{lemma}
\label{lem:cycle_with_tail}
The input graph $G$ is triangle free if and only if $G_8$ does not
have any $\mcC_7$ with a tail.
\end{lemma}
\begin{proof}
Observe that, if there exists a triangle $\tau$ in $G$, then in $G_8$,
there would be at least one $\mcC_7$ with a tail (in fact, the exact number 
would depend on the degree of the involved vertices). In the proof of~\Cref{lem:count_func}, we argued that each $7$-cycle in $G_7$ (which is
isomorphic to $G_8$) corresponds to a triangle in $G$. Also, by our
construction, if $G_8$ has a $\mcC_7$, then that $7$-cycle necessarily has a tail.
Therefore, existence of $\mcC_7$ with a tail in $G_8$ implies existence of a
triangle in $G$. This completes the proof of the lemma.
\end{proof}
\Cref{lem:degen,lem:count_func,lem:cycle_with_tail}  together prove the theorem: 
if there exists an algorithm $\mcA$ for the $\SUBC_{\mcC_k}$ problem with 
$T(\mcA) = O(mf(\degen))$, then $\mcA$ is an algorithm for the
$\TRIC$ problem (or the triangle detection problem in the case of $k=8$) 
with $T(\mcA) = O(m)$, where $T(\mcA)$ denotes the worst case time complexity of the algorithm $\mcA$.
\end{proof}

%%%%%%%%%%%%%%%%%%%%%%%%%%%%%%%%%%%%%%
%%%%%%%% FUTURE
%%%%%%%%%%%%%%%%%%%%%%%%%%%%%%%%%%%%%%
\section{Future Directions} \label{sec:future}
\begin{figure*}[h]
\centering
  \begin{tikzpicture}[nd/.style={scale=1,circle,draw,fill=blue,inner sep=2pt},minimum size = 6pt]    
    \matrix[column sep=1cm, row sep=0.2cm,ampersand replacement=\&]
    {
     \CountableSixVertexExample \\
 };
  \end{tikzpicture}
\caption{\Alg{counting} succeeds to count the number of distinct matches of $H$ in linear time for bounded (constant) degeneracy graphs. Each acyclic orientation of $H$ has a source vertex $s$, which is connected to exactly three vertices, as in $H\dir$. So, the largest $\drtss$ has at least four vertices (shown in green). Number of matches of the remaining vertices (shown in blue) could be counted using $\hashmap_2$}
\label{fig:6-vertex-countable}
\end{figure*}
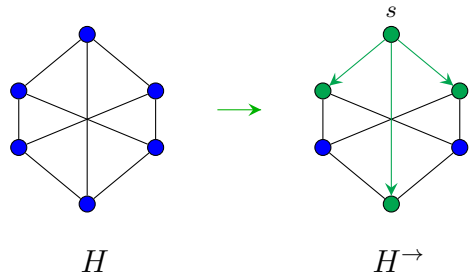

Although our algorithmic framework fails to produce a linear time algorithm for $\SUBC_{\mcC_6}$ in bounded degeneracy graphs, there are certain other 6-vertex subgraphs where it indeed succeeds. An easy example is $\SUBC_{\mcK_6}$. In fact, our framework gives a linear time algorithm for counting any constant size clique in bounded degeneracy graphs --- for each acyclic orientation of a clique, the source vertex construct a $\drtss$ covering all the remaining vertices.
There exists other non-clique 6-vertex subgraphs as well, where \Alg{counting} succeeds. Consider the subgraph $H$ shown in \Fig{6-vertex-countable}. It is easy to see that, any acyclic orientation of $H$ such as $H\dir$ has at least one source vertex $s$ that is a root of a $\drtss$ with four vertices.
%Also, notice that $H$ is not a star graph, so all the assumptions of \Alg{counting} are true.
Thus, we can solve $\SUBC_H$ in $O(m\degen^3)$ expected time.

Despite the chasm at six, there exist subgraphs $H$ with $6$-vertices 
(or more) such that $\SUBC_H$ admits a linear time algorithm in bounded degeneracy graph. We end this exposition with the following natural problem:

\vspace{0.2cm}
\emph{Characterize all subgraphs $H$ such that $\SUBC_H$ has a linear 
time algorithm in bounded degeneracy graphs.}

\section*{Acknowledgement}
We would like to thank David Helmbold for insightful discussions. In particular, David pointed out
that the lower bound for counting $8$-cycles does not follow 
from our construction, as we erroneously claimed in
the earlier version of this paper.

\bibliographystyle{plainurl}
\bibliography{refs}

\end{document}